%% file: main.tex
\documentclass[conference]{IEEEtran}
\IEEEoverridecommandlockouts
\usepackage{cite}
\usepackage{amsmath,amssymb,amsfonts}
\usepackage{graphicx}
\usepackage{textcomp}
\usepackage{xcolor}
\usepackage{algorithmicx}
\usepackage{algpseudocode}
\usepackage{algorithm}
\usepackage{xspace}
\usepackage{hyperref}
\usepackage{numprint}
\usepackage{todonotes}
\usepackage{booktabs}
\usepackage{balance}

\def\BibTeX{{\rm B\kern-.05em{\sc i\kern-.025em b}\kern-.08em
T\kern-.1667em\lower.7ex\hbox{E}\kern-.125emX}}
\include{macros}

\newcommand{\algo}[1]{\textsc{#1}}
\newcommand{\parents}[1]{\,\Pi_{#1}}
\newcommand{\children}[1]{\,child_{#1}}
\newcommand{\cluster}{\,\mathcal{S}}

\newcommand{\heft}{\algo{HEFT}\xspace}
\newcommand{\heftmm}{\algo{HEFTM-MM}\xspace}
\newcommand{\heftbl}{\algo{HEFTM-BL}\xspace}
\newcommand{\heftblc}{\algo{HEFTM-BLC}\xspace}

\newcommand{\MM}{M}

\newcommand{\rt}{rt}

\newcommand{\PD}{PD}

\newcommand{\new}[1]{{#1}}

\renewcommand{\iec}{i.e., }

\begin{document}

    \title{Memory-aware Adaptive Scheduling of Scientific Workflows on Heterogeneous Architectures\\
    }

\author{\IEEEauthorblockN{Svetlana Kulagina}
\IEEEauthorblockA{\textit{Humboldt-Universität zu Berlin, Germany} \\
ORCID: 0000-0002-2108-9425}
\and
\IEEEauthorblockN{Anne Benoit}
\IEEEauthorblockA{\textit{ENS Lyon \& IUF, France} \\
ORCID: 0000-0003-2910-3540
}
%email address or ORCID
\and
\IEEEauthorblockN{Henning Meyerhenke}
\IEEEauthorblockA{\textit{Humboldt-Universität zu Berlin, Germany} \\
ORCID: 0000-0002-7769-726X}
}

\maketitle

 \IEEEpubid{\begin{minipage}{\textwidth}\ \\[12pt]
 © 2025 IEEE. Personal use is permitted, but republication/redistribution requires IEEE permission.
 See https://www.ieee.org/publications/rights/index.html for more information.
 The published conference version of this paper appears in the Proceedings of 25th International Symposium on Cluster, Cloud and Internet Computing (CCGrid), IEEE Computer Society.
\end{minipage}}

    \begin{abstract}
The analysis of massive scientific data often happens in the form of workflows with
interdependent tasks. When such a scientific workflow needs to be scheduled on a parallel or distributed
system, one usually represents the workflow as a directed acyclic graph (DAG). 
The vertices of the DAG represent the tasks, while 
its edges model the dependencies between the tasks (usually data to be communicated to
successor tasks). When executed, each task requires a certain amount of memory and if that
exceeds the available memory, the execution fails.
The typical goal is to execute the workflow without failures (i.e., satisfying the memory
constraints) and with the shortest possible execution time (i.e., to minimize its makespan).

To address this problem, we investigate the memory-aware scheduling of DAG-shaped workflows on
heterogeneous platforms, where each processor can have a different speed and a different memory size.
We propose a variant of HEFT (Heterogeneous Earliest Finish Time) that, in contrast to the original, accounts for memory and
includes eviction strategies for cases when it might be beneficial to remove some data from memory
in order to have enough memory to execute other tasks.
Furthermore, while HEFT assumes perfect knowledge of the execution time and memory usage
of each task, the actual values might differ upon execution. Thus, we propose an adaptive
scheduling strategy, where a schedule is recomputed when there has been a significant variation in terms
of execution time or memory.
The scheduler has been closely integrated with a runtime system, allowing us to perform a thorough
experimental evaluation on real-world workflows. The runtime system warns the scheduler when
the task parameters have changed, and a schedule can be recomputed on the fly. The memory-aware
strategy allows us to schedule task graphs that would run out of memory with a state-of-the-art
scheduler, and the adaptive setting allows us to significantly reduce the makespan.

%                when tentatively assigning tasks, in order to
%        Its first step is to compute the weights of the tasks.
%        We suggest three variants: bottom levels as weights, bottom levels with impact of incoming edge weight,
%        and weights along the optimal memory traversal.
%        In the second step, we try assigning each task to each processors and execute the assignment that
%        is feasible with regard to memory size and gives the earliest finishing time to the task.
%        Sometimes, data corresponding to edge weights that is stored in the memory needs to be evicted in order to
%        assign a task to a processor.
%        We suggest two eviction strategies - largest files first and smallest files first.

%        Our experimental evaluation on real-world workflows and simulated (\skug{generated? They are generated from real-world wfs})
%        ones with real task and edge weights with up to 30,000 tasks shows that
%        respecting memory constraints only costs $11\%$ of runtime in comparison to a non memory-aware baseline.
%        Calculating task weights with the impact of memory gives a $x\%$ better makespans on a normal and $y\%$ better makespans
%        on small one.
%        Calculating task weights along the optimal memory traversal gives on average $z\%$ worse makespans, but improves
%        average memory utlization by $t\%$.

\end{abstract}

    \begin{IEEEkeywords}
        DAG, Heterogeneous platform, Adaptive \new{workflow} scheduling, Memory constraint.
    \end{IEEEkeywords}

\section{Introduction} %: \skug{Full: 0, Polished: 0}}
    \IEEEpubidadjcol
%%% CONTEXT %%%
The analysis of massive datasets, originating from fields such as genomics, 
remote sensing, or biomedical imaging -- to name just a few -- has become ubiquitous in science;
this often takes the form of workflows, \iec separate software components chained together
in some kind of complex pipeline~\cite{DBLP:journals/dbsk/LeserHDEGHKKKKK21}.
These workflows are usually represented as directed acyclic graphs (DAGs).
The DAG vertices represent the software components (or, more generally, the workflow \emph{tasks}),
while the edges model I/O dependencies between the tasks~\cite{adhikari2019survey,liu2018survey}.
Large workflows with resource-intensive tasks can easily exceed the capabilities of a 
single computer and are therefore executed on a parallel or distributed platform.
An efficient execution of the workflows on such platforms requires the mapping of tasks
to specific processors.
\new{Moreover, a task schedule -- \iec a valid execution order that respects the dependencies --
and possibly also starting times for the tasks are needed to increase utilization by reusing 
processors that have become idle after finishing a task.}
    \IEEEpubidadjcol

%%% MOTIVATION %%%
Modern parallel and in particular distributed computing platforms are often heterogeneous,
meaning they feature varying CPU speeds and memory sizes. 
In general, having different memory sizes per CPUs makes it more challenging for an algorithm 
to compute a schedule that respects all memory constraints -- meaning that no task is executed on a 
processor with less memory than needed for the task. Violating a memory constraint is, however, 
very important to avoid possibly expensive runtime failures and to provide a satisfactory user experience.
Hence, building on previous related %\hmey{You may want to add other refs not from us} 
work~\cite{gou2020partitioning,He21,DBLP:conf/icpp/KulaginaMB24}, we consider a scheduling problem 
formulation that takes memory sizes as explicit constraints into account. 
Its objective is the very common \emph{makespan}~\cite{liu2018survey}, 
which acts as proxy for the total execution time of a workflow.
However, to the best of our knowledge, the only memory-aware heuristics that would account for 
memory constraints partition the DAG and do not reuse processors once they have processed a part
of the graph. This approach leads to high values of makespan compared to a more fine-grained solution
with processor reuse. 

While previous work with memory constraints has focused on partitioning the DAG and not on 
reusing processors during execution, a seminal list scheduling heuristic for workflows on 
heterogeneous platforms, without accounting for the memory constraint, is HEFT 
(heterogeneous earliest finish time)~\cite{topcuoglu2002performance}.
It has two phases: (i) each task is assigned a priority and (ii) the tasks in a priority-ordered list are assigned
to processors, where the ``ready'' task with the highest priority is scheduled next on the processor
where it would complete its execution first. 
HEFT has been extended (e.g., by Shi and Dongarra~\cite{SHI2006665}) and adjusted 
for a variety of different scheduling problem formulations. 
Yet, none of them adhere to memory constraints as addressed in this paper -- 
see the discussion of related work in Section~\ref{sec:related-work}. 
%    (\skug{check in related work if true!}).
%        \skug{Note: 2 papers that deal with memory sizes, but model is very different!}
%    
Another limitation of HEFT (and many other scheduling strategies) in practice is their 
assumption that the task running times provided to them are accurate. In practice, this is 
not the case and deviations from user estimates or historical measurements are 
very common~\cite{hirales2012multiple}. As a consequence, it is advisable to adapt the schedule when \emph{major}
deviations occur. However, the original list-based schedulers, such as HEFT, are designed
for a static setting with accurate task parameters. 
%    
%    List-based schedulers such as HEFT are, however, not designed for 
%    such an adaptation~\cite{TODO}.\hmey{Svetlana, Anne: is this a fair statement? Please add ref (if any)}
%    \AB{Actually, we adapt HEFT as well, I'll reformulate...}
%%    and would compute a completely new schedule from scratch.

%%% CONTRIBUTION %%%
%\paragraph*{Contribution} 
The main contributions of this paper are both algorithmic and experimental:
\begin{itemize}
\item We formalize the problem with memory constraints, where communication buffers
are used to evict data from memory if it will be later used by another processor. 
\item We design three HEFT-based heuristics
that adhere to memory size constraints: \heftbl, \heftblc, and \heftmm.
M behind HEFT stands for \underline{m}emory, BL for \underline{b}ottom \underline{l}evel,
BLC for \underline{b}ottom \underline{l}evel with \underline{c}ommunication, 
and MM for \underline{m}inimum \underline{m}emory traversal.
The difference between the new heuristics is the way they prioritize tasks for processor assignment.

\item We implement a runtime system able to provide some feedback to the scheduler
when task requirements (in terms of execution time and/or memory) differ from the initial predictions, 
and we recompute a schedule, based on the reported deviations. 

\item We perform extensive simulations, first in the static case by comparing the schedules produced 
by these heuristics with the classical HEFT as baseline (the latter does not take memory sizes into account); 
while HEFT returns invalid schedules that exceed the processor memories and cannot execute correctly,
the new heuristics are able to successfully schedule large workflows and do so with reasonable makespans.

\item In the dynamic setting, we use a runtime system that allows us to simulate workflow executions.
The runtime system introduces deviations in running times and task memory requirements and communicates
them to the scheduler; the scheduler can then recompute a schedule. Without these recomputations,
most schedules become invalid after deviations, since the memory constraint is exceeded 
for most workflows, which demonstrates the necessity of a dynamic adjustment of the schedule. 

%    \hmey{Need to clarify: why is this SotA?}

%  \begin{itemize}
%      \item Static: we find that our heuristics are able to schedule all workflows correctly, and produce makespans similar to the baseline.
%      \item Adaptive: runtime system built, simulates workflow executions and deviations in running times and mem requirements of tasks
%      \item Answering requests of the runtime system for adaptation, the scheduler computes an improved schedule based on the reported deviations.
\end{itemize}

We first review related work in Section~\ref{sec:related-work}. Then, we formalize the model in Section~\ref{sec:model}
and present our algorithms in Section~\ref{sec:heuristics}. The adaptation of the heuristics in a dynamic setting is discussed in Section~\ref{sec:dyn}, and the results of our experiments are presented in Section~\ref{sec:expe}. Finally, we conclude
and provide future working directions in Section~\ref{sec:conc}.

\section{Related work} %: \skug{Full: 5, Polished: 4}}
\label{sec:related-work}

\subsection{HEFT-based algorithms}
%\subsection{Static list schedulers, especially HEFT-based algorithms}
%\label{sub:static-list-schedulers}
%
Introduced in 2002, HEFT~\cite{topcuoglu2002performance} is a list-based heuristic, consisting 
%HEFT and its successors consist 
of two phases: task prio\-ri\-tization/ordering and task assignment.
In the first phase, the algorithms compute bottom levels of the tasks based on some priorities (create the list),
and then schedule tasks in the order of these priorities.
The modifications of HEFT revolve around the way the priorities of the tasks are computed and the logic of the processor assignment.
All such algorithms assume a heterogeneous execution environment.

%During the task prioritization phase in~\cite{sulaiman2021hybrid}, the standard deviation of the computation cost % (between processors) 
%is computed and added to the mean value to account for differences between processor speeds.
%% TODO: Why ``between processors''? And what exactly is ``computation cost'' in the first place?
%In the processor choice phase, the entry task and the longest parent tasks are duplicated 
%during idle times on the processor.
%% TODO: Context of previous sentence unclear. Why is this done? What is the difference to others / our work in this respect?

%    Ref.~\cite{alebrahim2017task} computes the bottom level based on the difference of execution times on
%    the fastest and the slowest processors, divided by the speed ratio of these two processors.
%    When doing processor selection, the authors differentiate between the lowest execution time and earliest finishing time.
%    They choose the processor with the lowest execution time and cross over to other processors sometimes.
%    They build upon~\cite{shetti2013optimization}.\hmey{Last sentence ``hangs in the air''. Either drop or connect it properly.}

Some variants have been designed with various ways of ordering tasks, for instance based
on an optimistic cost table in 
PEFT (Predict earliest finish time)~\cite{arabnejad2014list}, or by combining the standard deviation
with the communication cost weight on the tasks in HSIP (Heterogeneous Scheduling with Improved task Priorities)~\cite{wang2016hsip}. 
% is a HEFT variant that computes an Optimistic
%Cost Table (OCT).
%The OCT is computed per task-processor pair and stores the longest shortest path from this task to the target task if this
%processor is chosen for this task.
%% TODO: is ``target task'' defined/clear?
%Ranking is based on OCT values.
%The processor choice stage minimizes the optimistic EFT, which is EFT plus the longest path to the exit node for each task. % TODO: is ``exit node'' == ``target task''? 
%
%The HSIP (Heterogeneous Scheduling with Improved task Priorities)~\cite{wang2016hsip} has an improved first step in
%comparison to HEFT.
%It combines the standard deviation with the communication cost weight on the tasks.
%In the second stage, the algorithm duplicates the entry task if there is a need for it.
%% TODO: Sounds like Ref. [32] above. What is the difference? (Why) Is it an improvement? Do we need to describe both?

The TSHCS (Task Scheduling for Heterogeneous Computing Systems) algorithm~\cite{alebrahim2017task} 
improves on HEFT by adding randomized decisions to the second phase.
%The decision is whether the task be assigned to the processor with the lowest execution time or to the processor that
%produces the lowest finish time.
The SDC algorithm~\cite{SHI2006665} considers the percentage of feasible processors in addition to a task’s
average execution cost in its weight.
%The selected task is then assigned to a processor which minimizes its Adjusted Earliest Finish Time (AEFT),
%which additionally notes how large the communication between the current node and its children will be on 
%average when scheduled on the current processor.
HEFT  can also be adapted in cloud-oriented environments~\cite{samadi2018eheft} and even combined with reinforcement learning techniques~\cite{yano2022cqga}, but none of these variants of HEFT consider memory constraints, 
to the best of our knowledge. 

%\medskip
\subsection{
%\noindent{\bf 
Memory-aware scheduling algorithms}
%\label{sub:mem-aware-algs}
%
%Only memory-aware scheduling algorithms are designed to respect memory constraints.
%%Respecting processor memories adds a constraint to a scheduling problem.
%%Therefore, only specifically memory-targeted algorithms address this issue.
%Moreover, 
The way processor memories are represented in the model has a decisive impact on the way the constraint
is formulated and addressed in the algorithm.
Different models of memory available on processors and memory requirements of tasks have been presented.

Marchal~et~al.~\cite{marchal2018parallel} assume a memory model where each processor has an individual memory available.
\new{Workflow tasks themselves have no memory requirements for their computations,
but they have input and output files that need to be stored in the memory.}
A polynomial-time algorithm for computing the peak memory needed for a parallel execution of such a workflow DAG is provided,
as well as an integer linear programming (ILP) solution to the scheduling problem.
The memory model \new{ includes no memory requirement of the task itself, only the weights of incoming and outgoing edges.
When the task starts, all input files are deleted  and all output files are added to memory.}

Other models consider for instance a 
%In an assumed 
dual-memory system~\cite{herrmann2014memory} where a processor can have access 
to two different kinds of memory, and each task can be executed on only one sort of memory.
%Communication happens only between these two kinds of processors (communication within 
%each group is ignored).
The authors then present an ILP-based solution for this problem formulation.
%
%The algorithm presented by Yao et al.
Yao et al.~\cite{yao2022memory} consider that each processor has its own internal memory, and all
processors share a common external one. The internal (local) memory is used to store the task files.
The external memory is used to store evicted files to make room for the execution of a task on a processor.
%All processors, including the original one, can access these files. 
%Each edge %in~\cite{yao2022memory} 
%has two weights -- the size of the files transferred along it,
%and the time of communication along this edge.
%The tasks themselves have no memory requirements, but need to hold all their incoming and outgoing files.
%
Ding \etal~\cite{ding2024ils}, in turn, consider connected processors with individual limited memories
forming a global memory, with different access times to memory and no weights on edges. An ILP
is proposed in this setting. 
%The collective set of memories forms the global memory to which each processor has access;
%however, the access time to global memory is different.
%Each memory access in the graph is modeled as a memory access token on the task, while the edges have no weights.
%The solved problem is how to allocate the initial input data in processor memories so that the overall
%execution is minimized and the memories are not exceeded.
%To this end, the authors propose an ILP model.
%%that minimizes the length of the critical path, including a greedy initial solution.
%
%In~\cite{rodriguez2019exploration}, the authors assume memory requirements on tasks represented as tiles.
%Each processor has individual memories to process the task, but only the shared memories store the tiles containing
%memory tiles occupied by memory tiles.
%
There are also some cloud-oriented models that include costs associated with memory usage~\cite{liang2020memory}.

Overall, there are a variety of memory models, but, to the best of our knowledge, the only study on a multiprocessor
platform that is fully heterogeneous, with individual memories, \new{is our own previous work~\cite{DBLP:conf/icpp/KulaginaMB24}.}
\new{Yet, in that paper, we only propose} a partitioning-based mapping of the workflow, without processor reuse.
\new{Since a processor is idle after a workflow partition has finished its execution 
in~\cite{DBLP:conf/icpp/KulaginaMB24},
there is no need for communication buffers to store data
that should be \new{exchanged} between processors when tasks are ready to execute.}

 \subsection{%
%\bigskip
%\noindent{\bf 
Dynamic/adaptive algorithms}
We finally review related work in a dynamic setting. With no variation in task parameters, 
    DVR HEFT~\cite{SANDOKJI2019482} rather considers that new tasks arrive in the system. 
    They use an almost unchanged HEFT algorithm in the static step, executing three slightly
    varying variants of task weighting and choosing the variant that gives the best overall makespan.
%    In the dynamic phase, they receive new tasks and schedule them on either idle processors or 
%    those processors that give them
%    the earliest finish time.
%    %Task failures are not covered.
%
%    Rahman~\etal 
    The dynamic critical path (DCP) algorithm for grids maps tasks to machines
    by calculating the critical path in the graph dynamically at every step~\cite{rahman2013}.
    %For all tasks they compute the earliest start time and absolute latest start time that are upper and lower bounds
    %on the start time of a task (differing by the slack this task has).
    %All tasks on this critical path have the same earliest and latest start times, because they cannot be delayed.
    The authors schedule the first task on the critical path to the best suitable processor and recompute the critical path.
    %The algorithm takes the first unscheduled task on the critical path each time and maps it on a processor identified for it.
    %If processors are heterogeneous, then the start times are computed with respect for the processor, and the minimum
    %execution time for the task is chosen.
    The heuristic also uses the same processor to schedule parent and children tasks, as to avoid data transfer between processors.
%    The approach is evaluated on random workflows of the size up to 300 tasks.

    Garg~\etal~\cite{GARG2015256} propose a dynamic scheduling algorithm for heterogeneous grids based on rescheduling.
    The procedure involves building a first (static) schedule with HEFT, periodic resource monitoring, 
    and rescheduling the remaining tasks. 
%    The resource model contains resource groups (small tightly-connected sub-clusters), connected between each other.
%    For each resource group, there is an own scheduler, and an overall global scheduler responsible for distributing
%    tasks to groups.
%    The static heuristic is HEFT with earliest start time as priority.
    Upon rescheduling, a new mapping is computed from scratch, 
    and this mapping is accepted if the resulting makespan
    is smaller than the previous one.
    \new{Only small workflows with less than 100 tasks are considered, hence it is possible to do
    these repeated computations from scratch in a reasonable time. }

De Olivera~\etal~\cite{de2012provenance} propose a tri-criteria (makespan, reliability, cost) \new{cost model and an adaptive
 scheduling algorithm for clouds (scheduling virtual machines).
The greedy algorithm based on this cost model  works in four steps, choosing the best resource types to execute the virtual
machine on, producing new cloud activities, setting up the granularity factor, and finally adapting the amount of resources
to fit the budget.}
The authors test four scenarios -- one preferring each criterion in the cost model and a balanced one.
The authors use workflows with less than ten tasks, but repeat them so that the execution has up to 200 tasks.
%They do not report the runtime of the scheduling algorithm, only the speedup and cost saving it produces.
%   The authors use provenance data to make scheduling decisions.

Daniels \etal~\cite{daniels1995robust} formalize the concept of robust scheduling with variable processing times
on a single machine.
The changes in task running times are not due to changing machine properties, but are rather task-related
and thus unrelated to each other.
The authors search for an optimal schedule
%    formulate a decision space of all permutations of $n$ jobs, and the optimal schedule 
in relation to a performance measure. % $\phi$.
Then, they proceed to formulate the Absolute Deviation Robust Scheduling Problem \new{with} a set of linear constraints.

While several related works consider building a new schedule once some variation has been observed,
we are not aware of work implementing a real runtime system that interacts with the scheduler
and has been tested on workflows with thousands of tasks, as we propose in this paper. Furthermore, 
we are not aware of any previous work discussing dynamic algorithms combined with memory constraints.

\section{Model} %: \skug{Full: 4, polished: 3}}
\label{sec:model}
The applications we target, large scientific workflows for which we do not have exact a priori knowledge,
are described in Section~\ref{sec.mod.work}. Then, the type of heterogeneous system on which
the applications are to be executed is presented in Section~\ref{sec.mod.plat}. The optimization problem
is defined in Section~\ref{sec.mod.pb}, and the key notation is summarized in Table~\ref{tabnotation}.

\subsection{Applications: Large scientific workflows}
\label{sec.mod.work}
Following common practice, we represent a workflow by a DAG (Directed Acyclic Graph) 
% The target applications, corresponding to large scientific workflows, are represented with a DAG (Directed Acyclic Graph) 
$G=(V,E)$. The set of vertices~$V$ corresponds to tasks, while edges express the
precedence constraints. Hence, a directed edge $e=(u,v)\in E$ means that task $u\in V$ must be executed 
before task~$v\in V$. A cost~$c_{u,v}$ is associated with each edge, representing the size of the 
output of task~$u$, to be used by task~$v$. 
Furthermore, $w_u$ is the number of operations performed by task~$u\in V$, 
and $m_u$ is the amount of memory required by task~$u$ to be executed. 
%    A workflow is modeled as a directed acyclic graph $G=(V, E)$, where $V$ is the set of vertices (tasks), and
%    $E$ is a set of directed edges of the form $e=(u,v)$, with $u,v\in V$, expressing precedence constraints between tasks.
%    Each task~$u \in V$  is performing $w_u$ operations, and it also
%    requires some amount of memory to be executed, denoted as~$m_u$.
%    For an edge $(u,v) \in E$, the weight~$c_{u,v}$  corresponds to the size of the output file 
%    written by task~$u$ and used as input by task~$v$.
We denote by $\parents{u}$ the tasks preceding task~$u\in V$, also called {\em parents}, 
which must be completed before $u$ can be started:   
%    The parents of a task~$u\in V$ are the directly preceding tasks that must be completed before $u$ can be started, i.e., the set of parents is
$ \parents{u} = \{v \in V: (v,u) \in E\}$. A {\em source} task is a task without parents. % is called a {\it source task}.
Similarly, the children of task $u\in V$ are %   The children tasks of~$u$ are the tasks following~$u$ directly according to the precedence constraints, i.e.,
$ \children{u} = \{v \in V: (u,v) \in E\}$, and a {\em target} task has no children. 
%A task without children is called a {\it target task}.
Each task may have multiple parents and children.

Note that $m_u$ is the total memory usage
of a task during its execution, including input and output files currently being read and written,
and hence the total memory requirement for executing task~$u$, denoted by~$r_u$,  is the maximum of the following:
(i) the total size of the files to be received from the parents, (ii) the total size of the files
to be sent to the children, and (iii) the total memory size~$m_u$ (which often reaches the maximum):
\[
    r_u = \max\left\{m_u , \sum_{v:(v,u)\in E}c_{v,u}, \sum_{v:(u,v)\in E} c_{u,v}\right\}.
\]

Furthermore, we operate in a context where we do not have perfect knowledge
of the task parameters ($w_u$ and $m_u$) before the tasks start their execution,
but only estimates~\cite{rahman2013,GARG2015256}.  
%. \AB{Add motivation: related work with variable task durations for instance...}
Hence, scheduling decisions are made on these estimated parameters, and
may be reconsidered at runtime when a task starts its execution and we know its exact parameters.

\subsection{Heterogeneous system}
\label{sec.mod.plat}
The target platform is a heterogeneous system $\cluster$ with $k$ processors 
%    The goal is to execute the workflow on a heterogeneous system, denoted as $\cluster$, which
%    consists of $k$ processors 
    $p_1, \dots, p_k$.
    For $1 \leq j \leq k$, each processor $p_j$  has an individual memory of size $M_j$, a communication
    buffer of size $MC_j$ and a speed~$s_j$.
    We can decide to evict some data from the main memory when we are sending the data
    to another processor; it then stays in the communication buffer until it has been sent.
    The execution time of a single task~$u\in V$ on a processor~$p_j$ is~$\frac{w_u}{s_j}$,
    and all 
    %We assume that all 
    processors are connected with an identical bandwidth~$\beta$.
%    \hmey{Maybe mention that variable bandwidths are part of future work...?}

    We keep track of the current ready time of each processor and each communication
    channel, $\rt_j$ and $\rt_{j,j'}$, for each processor $j$ and all pairs~$(j,j')$.
    Initially, all the ready times are set to~$0$.
    We also keep track of the currently available memory, $availM_j$ and $availC_j$,
    on the processor memory and communication buffer, respectively.
    Furthermore, $\PD_j$ is a priority queue with the {\em pending data}
    that are in the memory of size $\MM_j$ but may be evicted to be communicated if
    more memory is needed on~$p_j$. They are ordered by non-decreasing size and
    correspond to some $c_{u,v}$.

In order to compute the memory requirement of a DAG, we use \algo{memDag}~\cite{KAYAASLAN20181},
an algorithm that 
%    We use the \algo{memDag} algorithm developed by Kayaaslan \etal~\cite{KAYAASLAN20181} to compute
%    the memory requirement; it 
    transforms the workflow into a series-parallel graph
    and then finds the traversal that leads to the minimum memory consumption.

    \begin{table}
        \begin{center}
            \begin{tabular}{rl}
                \hline
                \textbf{Symbol}                       & \textbf{Meaning}                                         \\
                \hline
                $G = (V, E)$                          & Application DAG (tasks and edges)       \\
                $\parents{u}$, $\children{u}$         & Parents of task $u$, children of task $u$            \\
                $w_u$                                 & Number of operations of task $u$         \\
                $c_{u,v}$                             & Size of output file for edge $(u,v)\in E$         \\
                $m_u$                                 & Amount of memory required by task $u$                                \\
                $r_u$ 				& Total memory requirement for task~$u$ \\
%                $F$, $\mathcal{F}$                    & A partitioning function and the partition it creates     \\
%                $V_i$                                 & Block number $i$                                         \\ %\wrt~some $F$   \\
                $\cluster$, $k$                    & Computing platform, total number of processors           \\
%                $p_j$, proc($V_i$)                          & Processor number $j$, processor of block $V_i$                 \\
                $M_j$, $MC_j$, $s_j$                               & Memory size, comm. buffer size, and speed of proc.\ $p_j$                          \\
                $\beta$                     & Bandwidth in the compute system                                \\
                $bl(u)$                      & Bottom level of task $u$ \\
                $ST(u,p_j)$	 & Start time of task~$u$ on~$p_j$ \\
               $FT(u,p_j)$	 & Finish time of task~$u$ on~$p_j$ \\
%                $\mu_G$, $\mu_i$ & Makespan of the entire workflow $G$ and of a block $V_i$               \\
%                $\Gamma = (\mathcal{V}, \mathcal{E})$                      & Quotient graph, its vertices and its edges        \\
%                $r_u$, $r_{V_i}$                            & Memory requirement of task $u$ and of block $V_i$                 \\
                \hline
            \end{tabular}
        \end{center}
        \caption{Key Notation} \label{tabnotation}
    \end{table}

\subsection{Optimization problem}
\label{sec.mod.pb}

The goal is to find a schedule of the DAG~$G$ for the $k$ processors,
so that the makespan (total execution time) is minimized while
respecting memory constraints. If a processor runs out of memory to execute
a task mapped on it, the schedule is said to be {\em invalid}.

Since tasks are subject to variability, we aim at minimizing the actual makespan
achieved at the end of the execution, while decisions may be taken \wrt 
the estimated task parameters.

Note that the problem is already NP-hard even in the homogeneous case and 
without memory constraints, because of the DAG structure of the application. 
Hence, we focus on the design of efficient scheduling heuristics. 

\section{Scheduling heuristics}
\label{sec:heuristics}
We design variants of HEFT that account for memory usage and aim at minimizing the makespan.
First, we present in Section~\ref{sec.heft} the baseline HEFT heuristic; as it does not account for the memory,
it may return invalid schedules that will not be able to run successfully on the platform (due to 
running out of memory).  Then, Section~\ref{sec.heftm} focuses on the presentation of the novel
heuristics, including eviction strategies to move some data in communication buffers
in case there is not enough memory available on some processors.

\subsection{Baseline: original HEFT without memories}
\label{sec.heft}
Original HEFT does not account for memory sizes.
Its schedules can be invalid if tasks are assigned to processors without enough memory.
These solutions can be viewed, however, as a ``lower bound'' for a valid solution that 
respects memory constraints.

HEFT works in two stages.
In the first stage, it computes the ranks of tasks by computing their non-increasing bottom levels.
The bottom level of a task is defined as
\[
bl(u) = w_u + \max_{(u,v)\in E} \{c_{u,v} + bl(v)\}
\]
(with the max yielding $0$ if there is no outgoing edge).
The tasks are sorted by non-decreasing ranks.

In the second stage, the algorithm iterates over the ranks and tries to assign the task to the processor where it
has the earliest finish time.
We tentatively assign each task~$v$ to each processor~$p_j$.
The task's starting time $ST(v,p_j)$ on processor~$p_j$ is dictated by the maximum between the ready time of the processor~$rt_j$
and all communications that
must be orchestrated from predecessor tasks $u\notin T(p_j)$.
The starting time is then:

{\footnotesize{   \[ST(v, p_j) = \max{ \Large\{rt_j, \max_{ u \in \Pi(v)}\{ FT(u)+ \frac{c_{u,v}}{\beta} , 
rt_{proc(u), p_j} + \frac{c_{u,v}}{\beta}  \} \Large\} } . \]}}

Finally, its finish time on $p_j$ is
$FT(v,p_j) = st_v + \frac{w_v}{s_j}$.

Once we have computed all finish times for task~$v$,
we keep the minimum $FT(v,p_j)$ and assign task~$v$
to processor~$p_j$.

\textit{Assignment to processor. }
When assigning the task, we set the ready time $rt_j$ of  processor~$j$ to be the finish time of the task.
For every predecessor~$u$ of task~$v$ that has been assigned to another processor~$j'$, we adjust ready times on
communication buffers $rt_{j', j}$: % for every predecessor $u$'s processor $j'$: 
we increase them by the
communication time $c( u,v) / \beta$.

\subsection{Memory-aware heuristics}
\label{sec.heftm}
Like the original HEFT, the memory-aware versions of HEFT consist of two stages:
first, they compute the task ranks,
and second, they assign tasks to processors in the order defined in the first stage.
We consider three variants of HEFT accounting for memory usage (HEFTM), which only
differ in the order they consider tasks to be scheduled in the first stage.

\medskip
\noindent{\bf Compute task ranks. }

Our three variants of memory-aware HEFT work as follows:
\begin{itemize}
\item    
  HEFTM-BL orders tasks by non-increasing bottom levels, where the bottom
  level is defined as
  $$bl(u) = w_u + \max_{(u,v)\in E} \{c_{u,v} + bl(v)\}$$
  (max yields $0$ if there is no outgoing edge).

\item    
  HEFTM-BLC %: from the study of the fork (see below), it seems important
  %  to also account for the size of the data as input of a task,
  gives more priority to tasks with potential large incoming communications,
  hence aiming at clearing the memory used by files as soon as possible,
  to have more free memory for remaining tasks to be executed on the processor.
  Therefore, for each task, we compute a modified bottom level accounting for communications:
  $$blc(u) = w_u + \max_{(u,w)\in E} \{c_{u,w} + blc(w)\} + \max_{(v,u)\in E} c_{v,u}   . $$

%    \skug{avoid having mixed ranks, when the memory size of the lower task is not taken into account}

\item   
  Finally, HEFTM-MM orders tasks  in the order returned by %as dictated by MinMem.
  the  \algo{memDag} algorithm~\cite{KAYAASLAN20181}, which corresponds to a traversal
  of the graph that minimizes peak memory usage.
\end{itemize}

\bigskip
\noindent  {\bf Task assignment. }

Then, the idea is to pick the next free task in the given order,
and greedily assign it to a processor, by trying all possible options
and keeping the most promising one. We first detail how a task
is tentatively assigned to a processor, by carefully accounting for the memory usage.
Next, we explain the steps to be taken to effectively assign a task to a given processor.

\medskip
\noindent{\em Tentative assignment of task~$v$ on $p_j$.}\\
{\bf Step 1.} First, we need to check that for all predecessors~$u$ of~$v$ that are mapped
on~$p_j$, the data $c_{u,v}$ is still in the memory of~$p_j$,
i.e., $c_{u,v}\in PD_j$. Otherwise, the finish time is set to~$+\infty$ (invalid choice).

\smallskip
\noindent{\bf Step 2.} Next, we check the memory constraint on~$p_j$, by computing
$$Res = availM_j - m_v - \!\!\!\! \sum_{u \in \Pi(v), u\notin T(p_j)}  \!\!\!\!\!\!\!\!\{c_{u,v}\}
- \sum_{w\in Succ(v)} \!\!\!\!\!\! \{c_{v,w}\}.$$

$T(p_j)$ is the set of tasks already scheduled on $p_j$; by Step 1, their files are
already in the memory of~$p_j$. However, the files from the
other predecessor tasks must be loaded into memory before executing task~$v$,
as well as the task data of size $m_v$, and the data generated for all successor tasks.
$Res$ then checks whether there is enough memory; if it is negative,
we have exceeded the memory of~$p_j$ with this tentative assignment.
In this case ($Res <0$), we try to evict some data from memory so that there is enough memory to execute task~$v$.
Clearly, we need to evict at least $Res$ data.
To this end, we propose a greedy approach, which evicts the largest files of $\PD_j$ until data of size $Res$ have been evicted. 
A variant where the smallest files are evicted first has been tested; it led to comparable results. 
%    
%    in order to avoid costly communications.
%    \AB{FYI We initially discussed evicting the largest files, but this leads to
%    large communications and does not seem efficient after all... Maybe we can think of another
%    approach that would take into account both data size and bottom level...}
While tentatively evicting files, we remove them from the list of pending memories and move them into a list
of memories pending in the communication buffer.
We keep track of the available buffer size, too -- every time a file is moved into the pending buffer, 
the available buffer size is reduced by the file size.

If, after tentatively evicting all files from $\PD_j$, we still do not have enough memory, or if we exceed the size of the available buffer during this process, we set the finish time to $+\infty$ (indicating an invalid choice).

\smallskip
\noindent{\bf Step 3.} We tentatively assign task~$v$ on $p_j$.
Its starting time $ST(v, p_j)$ on $p_j$ is dictated by the maximum between $rt_j$ and all communications that
must be orchestrated from predecessor tasks $u\notin T(p_j)$.
The starting time is therefore:\\[-.7cm]

{\small{ \begin{align*}
ST(v, p_j) & = \max  \large\{rt_j,  \\
& \max_{ u \in \parent(v), u\notin T(p_j)}\{ FT(u) , rt_{proc(u), p_j}\} + \frac{c_{u,v}}{\beta} \Large\} . 
\end{align*}
}}

\noindent  Finally, its finish time on $p_j$ is 
$FT(v,p_j) = ST(v, p_j) + \frac{w_v}{s_j}$.

\medskip
\noindent{\em Assignment of task~$v$.}\\
Once we have computed all finish times for task~$v$,
we keep the minimum $FT(v,p_j)$ and assign~$v$
to processor~$p_j$. 
In detail:
\begin{itemize}
\item 
  We evict the files corresponding to edge weights that need to be evicted to free the memory.
  We remove these files from pending memories
  $PD_j$, add them to pending data in the communication buffer, and reduce the available buffer size accordingly.
\item 
  We calculate the new $availM_j$ on the processor.
  Next, we subtract the weights of all incoming files from predecessors assigned to the same processor
  and add the weights of outgoing files generated by the currently assigned task.
  
\item  
  For every predecessor of~$v$ that has been assigned to another processor, we adjust ready times on
  communication buffers $rt_{j', j}$ for the processor~$j'$ that the predecessor $u$ has been assigned to: we increase them by the
  communication time $c( u,v) / \beta$.
  We also remove the incoming files from either the pending memories or pending data in buffers of these other
  processors, and increase the available memory or buffer sizes on these processors.
  
\item 
  We compute the correct amount of available memory for~$p_j$ (for when the task is done).
  Then, for each predecessor that is mapped to the same processor, 
  we remove the pending memory corresponding to the weight of
  the incoming edge, also freeing the same amount of available memory (increasing $availM_j$).
  For each successor, we rather add the edge weights to pending memories and reduce $availM_j$ 
  by the corresponding amount.
\end{itemize}

%    \subsection{The fork}
%    We look at the behavior of these heuristics on a fork graph,
%    where there is a root task~$T_0$, producing $n$ files $f_1, \ldots, f_n$
%    to be used by tasks $T_1, \ldots, T_n$ ($f_i = c_{0,i}$).
%
%    Without memory, this problem is NP-complete; this is equivalent
%    to 2-partition if the tasks have $w_i=a_i$, and all files are of size~$f_i=0$,
%    and with two processors. Half of the tasks must be sent to the processor
%    on which $T_0$ is not executed, and the optimal makespan is
%    $w_0+\frac{1}{2}\sum_{1\leq i \leq n} w_i$.
%
%    However, with an infinite number of identical processors, it can be
%    solved in polynomial time: sort tasks by non-decreasing $f_i+w_i$;
%    the $k$ tasks with smallest $f_i+w_i$ are then sent to another processor,
%    while the remaining $n-k$ tasks are executed locally (try all values of $k$).
%
%    With heterogeneous processors, it is probably NP-complete again
%    because we could ensure that there are only two processors fast enough
%    and get back to the 2-partition...
%
%    We also had an example where evicting large files first in step 2
%    can lead to arbitrarily bad makespan. Consider a fork with $n=2$,
%    $f_1=1$, $w_1=2$, $f_2=100$, $w_2=1$, and memory constraint
%    imposes that we free one unit of memory before executing one
%    of the tasks\ldots Actually the new version with BLC would start
%    considering $T_2$ and be fine in this case\ldots
%
%
%    \AB{Can we prove that we have (maybe) a 2-approximation,
%        at least for the fork? What worst-case can we think of? }

\section{Dynamic scenario}
\label{sec:dyn}
In a workflow execution environment, the scheduling method interacts with the runtime environment, which provides information such as resource estimates.
This information may include memory usage, running time, graph structures, or the status of the underlying infrastructure.
In order to ensure that the information is up to date, a monitoring system observes the workflow execution and collects metrics for tasks and the underlying infrastructure.
By incorporating dynamic monitoring values, e.g., the resources a task consumed, the runtime environment can incorporate the data into the prediction model to provide more accurate resource predictions.
Also the underlying infrastructure can change during the workflow execution.
Examples are processor failures, node recoveries, or acquisition of new nodes.
Even if the hardware infrastructure does not change, the set of nodes provided as a scheduling target might change due to release or occupation in shared cluster infrastructures.
As infrastructure information and resource predictions are dynamically updated and provided to the scheduler during workflow runtime, the previous schedule may become invalid, so that a new one must be calculated.

For state-of-the-art memory prediction methods, a cold-start median prediction error for heterogeneous infrastructures
of approximately 15\% has been observed~\cite{malik2013execution}.
Online prediction methods were able to significantly reduce the error during runtime, with the reduction reaching up to one third of the cold-start error~\cite{baderDiedrichDynamic2023,witt2019learning}.
%For instance, Nadeen~et~al.\cite{} report an error of 10\%, 11\%, and 15\% while the task prediction errors shows a normal and exponential distribution.
%Bader~et~al.~ report a prediction error between 13\% and 17\% for their method, showing an exponential task error distribution.
% @Svetlana, willst du sowas für deine Experimente? Also die Daten, welche du dann konfigurieren kannst?
Such a dynamic execution environment requires a dynamic scheduling method where the schedule can be recomputed during workflow execution.

\subsection*{Retracing the effects of change on an existing schedule}
After the monitoring system has reported changes, we need to assess their impact on the existing schedule.
These changes can invalidate the schedule (\eg if there is not enough memory for some tasks to execute anymore),
they can lead to a later finishing time (\eg if some tasks are longer and they delay other tasks), or they can have no effect (\eg if new processors
joined the cluster, but the old schedule did not account for them).
To assess the impact, we need to retrace the schedule.

First, we find out if at least one processor that had assigned tasks has terminated operation -- this instantly invalidates the
entire schedule.
We then iterate over all tasks of the workflow in a topological order -- 
any of the orderings given by rankings BL, BLC or MM is a topological ordering.
We then repeat steps similar to those we did during tentative assignment in the heuristics, 
except that we do not choose a processor
anymore, but rather we check whether the current processor assigned to the task still fits.

For each task $v$, we first assess its current memory constraint $Res$ using Step 2 from the heuristic.
The factors that affect $Res$ are possible changes in $m_v$, in $c_{u,v}$ from predecessors $u$ 
or $c_{v,w}$ from successors $w$,
available memory $availM_j$ on the processor (due to either changed $M_j$ or changed memory requirements 
from other tasks).
If $Res$ was originally positive (no files evicted from memory into the communication buffer), 
then it has to stay this way -- otherwise, evicted files can invalidate subsequent tasks.
If $Res$ was originally negative, then we need to make sure that evicted files still fit into the communication buffer.
If either $Res$ is newly negative or the communication buffer is not large enough,
then this invalidates the schedule.
We update the $availM_j$ and $availMC_j$ values according to the new memory constraints.

Then, we can re-calculate the finish time of the task on its processor like in Step 3.
The factors that affect it are changes in own execution time $w_v$ of the tasks, changed ready time of the processor
(after delayed previous tasks), and changed communication buffer availability.

Finally, after having updated the processor's values, we move on to the next task.

%    \subsection{Approximation}
%    \hmey{Rough notes:}
%    Let's use a fork to see how the algorithm behaves and if it provides some approximation. Our current intuition is that, if the memory constraint is ignored, HEFTM-BLc provides a $2$-approximation (to be proved).

\section{Experimental evaluation}
\label{sec:expe}
We first describe the experimental setup in Section~\ref{sec:setup}. 
Then, we report results on static experiments to assess the performance
of the memory-aware heuristics in Section~\ref{sec.expe.static}, before
discussing the heuristics' behavior in a dynamic setting in Section~\ref{sec.expe.dyn}. 
Finally, we report running times of the heuristics in Section~\ref{sec.expe.t}. 

Note that we consider HEFT as the baseline; experiments to compare the novel heuristics
with the closest competitor~\cite{DBLP:conf/icpp/KulaginaMB24} obtained much better results 
(at least 10x smaller makespans). 
Indeed, we have much more flexibility in the current setting where processors are reused, since tasks
are not just partitioned. Hence, the comparison with~\cite{DBLP:conf/icpp/KulaginaMB24} seemed not fair
and we believe the natural baseline is HEFT in this setting. 
We are not aware of other dynamic scheduling algorithms that account for memory usage 
and hence could not find other competitors from the literature apart from HEFT.

\subsection{Experimental setup}
\label{sec:setup}
All algorithms are implemented in C++ and compiled with g++ (v.8.5.0).
The experiments are managed by simexpal~\cite{DBLP:journals/algorithms/AngrimanGLMNPT19} and executed on workstations with 192 GB RAM and 2x 12-Core Intel Xeon 6126 @3.2 GHz
and CentOS 8 as OS.
Code, input data, and experiment scripts are available to allow reproducibility of the results at~\url{https://zenodo.org/records/13919214}~and~\url{https://zenodo.org/records/13919302}.

Before presenting the results, we first describe the set of workflows we used for the evaluation and
then the clusters on which the workflows were scheduled.
We also explain the way the runtime system works and interacts with our scheduler. 

\medskip
\subsubsection{Workflow instances}
We run experiments on a set of several real-world workflows from Ref.~\cite{lotaru}
(atacseq, bacass, chipseq, eager, and methylseq). We use the WFGen generator~\cite{COLEMAN202216} to
create larger variants of the workflows above and add the instances created this way to the set.

\medskip

\paragraph{Workflow graphs}
For the five real-world workflows, their nextflow definition (see~\cite{ewels2020nf}) was downloaded from the
respective repository and transformed into .dot format using the nextflow option ``-with-dag''.
The resulting DAG contains many pseudo-tasks that are only internal representations in nextflow
(and not actual tasks); that is why we removed them.

For the size-increased workflows, the graph is produced by the WFGen generator, based on a {\em model workflow} and
the desired number of tasks.
We used the real-world workflows as models, except for bacass since it leads to errors in the generator.
As number of tasks, we use: 200, \numprint{1000}, \numprint{2000}, \numprint{4000}, \numprint{8000}, \numprint{10000},
\numprint{15000}, \numprint{18000}, \numprint{20000}, \numprint{25000}, and \numprint{30000}.
We divide the workflows into four groups by size: tiny ones with up to \numprint{200} tasks, small ones with \numprint{1000} to \numprint{8000} tasks,
middle ones with \numprint{10000} to \numprint{18000} tasks, and big ones with \numprint{20000} to \numprint{30000} tasks.

\new{Note that, due to a varying nature of the workflow generator, some workflows can have smaller makespans than their
smaller counterparts due to more parallelism at nodes with higher out-degree or other factors of the workflow's
internal composition.}

\medskip

\paragraph{Task and edge weights}
For the real-world workflows, we use historical data files provided by Bader~\etal~\cite{lotaru}.
The columns in these files are measured Linux PS stats, acquired during an execution of a nextflow workflow.
Each row corresponds to an execution of one task on one cluster node.
Since the operating system cannot distinguish between (a) the RAM the task uses for itself and (b) the RAM it uses
to store files that were sent or received from other tasks, the values in the historical data are total memory requirements (input/output files plus memory consumption of the computation).
In a similar manner, the historical data provided by Ref.~\cite{lotaru} do not store the actual weights of edges between tasks, but only the overall
size of files that the task sends to all its children.

For each task, historical data can contain multiple values, obtained from the runs with different input sizes.
The same workflow can require different memory capacities and take different times to execute
depending on the size of its input.
We simulate these various runs by obtaining values corresponding to each input size.
For each of the four families, there are five input sizes; we thus run each workflow in five variants corresponding to these inputs.

Not all tasks have historical runtime data stored in the tables.
In fact, for two workflows, Bader~\etal do not provide data for more than 50\% of the tasks.
For two more, around 40\% of the tasks have no historical runtime data stored.
Hence, in the absence of historical data about a task, we give it fixed weights:
an execution time of $1$, a memory requirement of $50 MB$, and files written and received of $1KB$.
These values align with the findings of Ref.~\cite{lotaru} about small tasks.

\medskip

\subsubsection{Target computing systems}
%    \skug{Because of the normalized nature of the task/edge weights, I will now be saying that our larger cluster has 192 GB and it
%    is represented as 192 000 (with 3 zeroes). For the memory-constrained cluster (192 00 with 2 zeroes) I will say that it has 10 times
%    less memory - 19.2 GB. Other processors will be dealt with similarly.}

To fully benefit from the historical data, the  {\em default} experimental environment
that we consider is a cluster based on the same six
kinds of real-world machines that were used in the experimental evaluation in Ref.~\cite{lotaru}.
We set the number of each kind of node to $12$, yielding 72 processors in total.  % the whole cluster.
Each machine has a (normalized) CPU speed and a memory size (in GB), and we list them as tuples (name, speed, memory):
($local$, 4, 16) -- very slow machines; ($A1$, 32, 32), ($A2$, 6, 64), ($N1$, 12, 16) -- average machines;
($N2$, 8, 8) -- machine with very small memory; and ($C2$, 32, 192) -- {\em luxury} machine with high speed and
large memory (see Table~\ref{tab:procs}).

\begin{table}[tb]
    \begin{center}
        \begin{tabular}{c|c|cc}
            \toprule
            Processor  %& Amount
            &  CPU speed   & \multicolumn{2}{c}{Memory size (GB)} \\
            name & (GHz) & {\em default} & {\em mem-constrained} \\
            \midrule
            local                    & 4                    & 16     & 1.6 \\
            A1                      & 32                   & 32     & 3.2 \\
            A2                      & 6                    & 64    & 6.4 \\
            N1                      & 12                   & 16     & 1.6 \\
            N2                      & 8                    & 8      & 0.8\\
            C2                      & 32                   & 192   &  19.2\\
            \bottomrule
        \end{tabular}
    \end{center}
    \caption{Cluster configurations.}
    \label{tab:procs}
\end{table}

We also consider a more constrained setting by varying the cluster configuration. 
The {\em memory-constrained cluster} consists of 72 nodes (12 of each kind) as the default cluster,
but each node has 10 times less memory. Hence,  
the {\em luxury} machine $C2$ has $19.2$~GB memory in this setting instead of $192$~GB, 
$A2$ has $6.4$~GB instead, of $64$~GB etc.
The processor speeds and their relations stay unchanged (see Table~\ref{tab:procs}).

Note that in both clusters, we set the size of the communication buffer to be equal
to ten times the memory size.

\medskip

\subsubsection{Runtime system}
\label{ss:runtime-sys}
To simulate the execution of a workflow, we implemented a runtime system.
It reads the historical data and builds weights for tasks as explained above.
%    These weights are then treated as the ``truth'' about the execution during the simulation.
In the static case, these values are being sent to the scheduler, which builds a schedule
according to these weights.
In the dynamic setting, in turn, the runtime system applies a deviation function to the values.
This function computes a normally distributed random deviation value, where the initial value
is the mean and the deviation is~$10\%$.
This scenario corresponds to the real-world scenarios identified in Ref.~\cite{lotaru} and
other works dedicated to predicting task running times~\cite{da2015online,da2013toward}.

Thus, the scheduler receives deviation values and makes decisions based on them\new{, which
leads to an execution mode  \textit{with recomputation}.}
This mode leads to several types of possible issues:
\begin{itemize}
\item 
  A processor is blocked by another task. If the scheduler underestimated the execution time of a task,
  it will block another one from starting.
\item 
  A predecessor has not finished yet.
  The scheduler may request a task to start on its processor, while
  some of the predecessors of the task have in fact not yet completed their execution -- 
  the task is therefore not yet ready.
%        A similar situation, but a task could start on its processor, but one or more of its parents a
%        are not ready yet.
\item 
  Not enough memory. If the scheduler underestimated the amount of memory a task requires, this task might not be able to execute on a chosen processor.
\item 
  A task took less time than expected. We only consider this case if a task took more than 10\% less
  time than expected. In this case, we want to exploit the newly acquired free time by possibly starting other tasks earlier.
\end{itemize}

\new{Alternatively, the runtime system can execute \textit{without recomputation}.
It then follows the original schedule without sending the deviation values to the scheduler.

In this scenario, the runtime system waits with task execution for a busy processor to become available if another task scheduled on it takes longer than expected.
If, however, the predecessor task is faster than expected, the runtime system leaves the processor of the current task temporarily idle.
Finally, if there is not enough memory, the schedule is declared invalid and the execution stops.

The comparison to the makespan of the execution without recomputation shows how much improvement the exploration of dynamic
changes brings.
Moreover, the number of schedules that manage to stay valid despite the changes is an indicator of their robustness and of the quality of the scheduler.}

\subsection{Results in a static setting}
\label{sec.expe.static}
We first study the heuristics' behavior when the weights do not change during runtime,
hence the scheduler has perfect knowledge of the task memory requirements and execution times.
We have compared two eviction strategies, starting with large files first or small files first,
and did not observe any significant changes in terms of schedule validity or makespan.
Thus, we only present results with the eviction of largest files. % first.
%TODO: Why ``first''? SK: don't see a problem here. We first kick the largest file, then the second largest etc
%as opposed to first the smallest, then second smallest ...

%    Our preliminary experiments have shown that evicting large or small files does not give a notable change in makespan.

\medskip
\subsubsection{Scheduling on the default cluster}
\begin{figure}[tb]
  \centering
  \vspace{-0.4cm}
  \includegraphics[width=1.0\columnwidth] {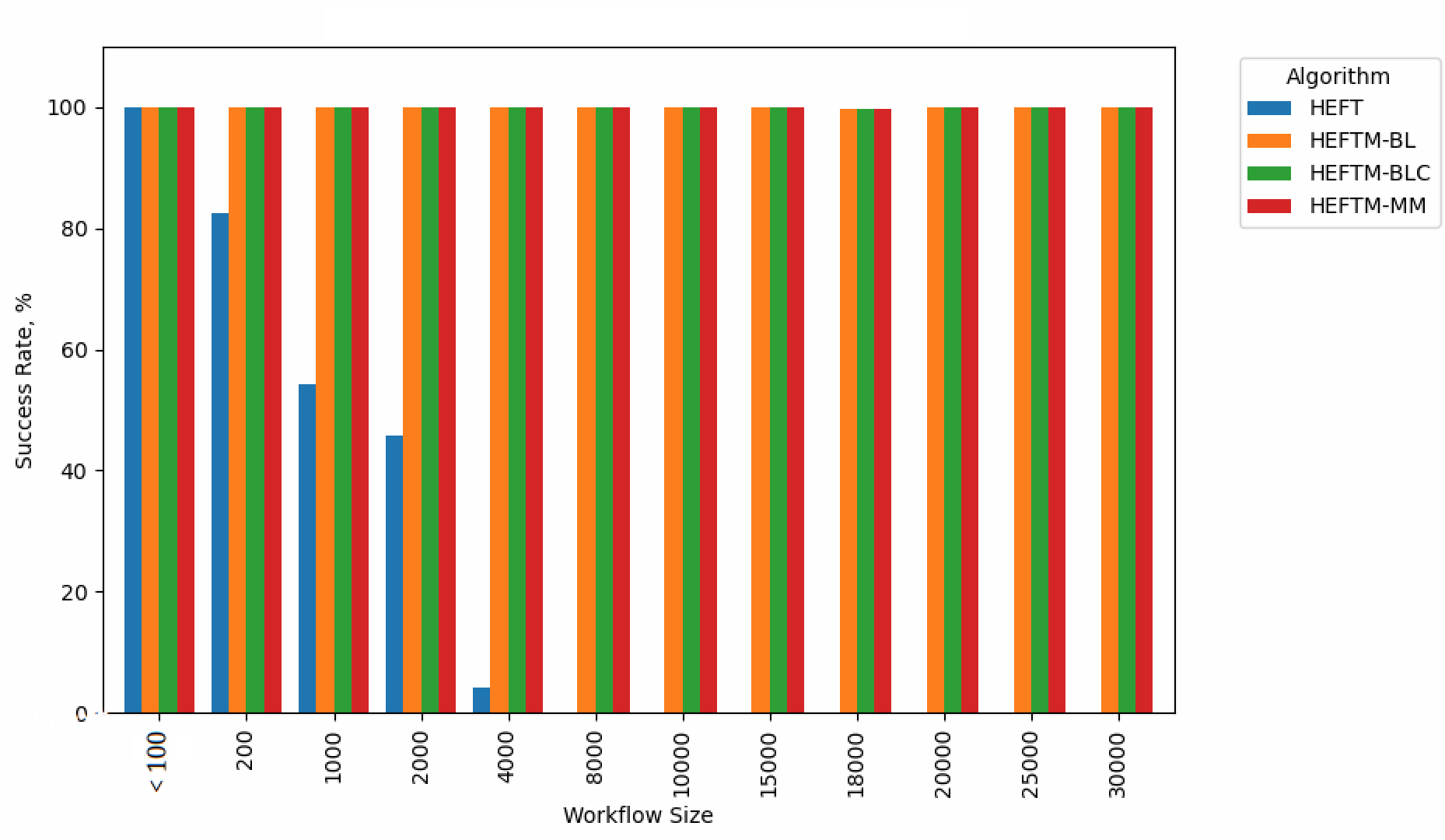}
  \caption{Success rates by workflow size and algorithm on the default cluster. Higher is better.}
  \label{fig:success-rates-large}
%  \vspace{-0.15cm}
\end{figure}

%
%    \paragraph{Normal cluster}
%
On the default cluster,  the three memory-aware heuristics are able to schedule all workflows
(see Figure~\ref{fig:success-rates-large}), while
the baseline \heft has a success rate of only $24.2\%$ ($75.7\%$ failure rate).
Indeed, \heft is only able to schedule small workflows; it cannot schedule any workflow 
with more than $4000$ tasks correctly as some tasks run out of memory. As soon as we are not in a setting
with abundant processing resources for small workflows, it is hence necessary to adopt
a memory-aware strategy in order to produce valid schedules.
%    Small workflows are less constrained by the memory requirement, so in the situation of abundant processing resources, even
%    a not memory-aware approach is able to produce valid schedules.
%    However, this advantage quickly vanishes as workflows grow larger.

\begin{figure}[tb]
    \centering
  \vspace{-0.34cm}
  \includegraphics[width=1\columnwidth] {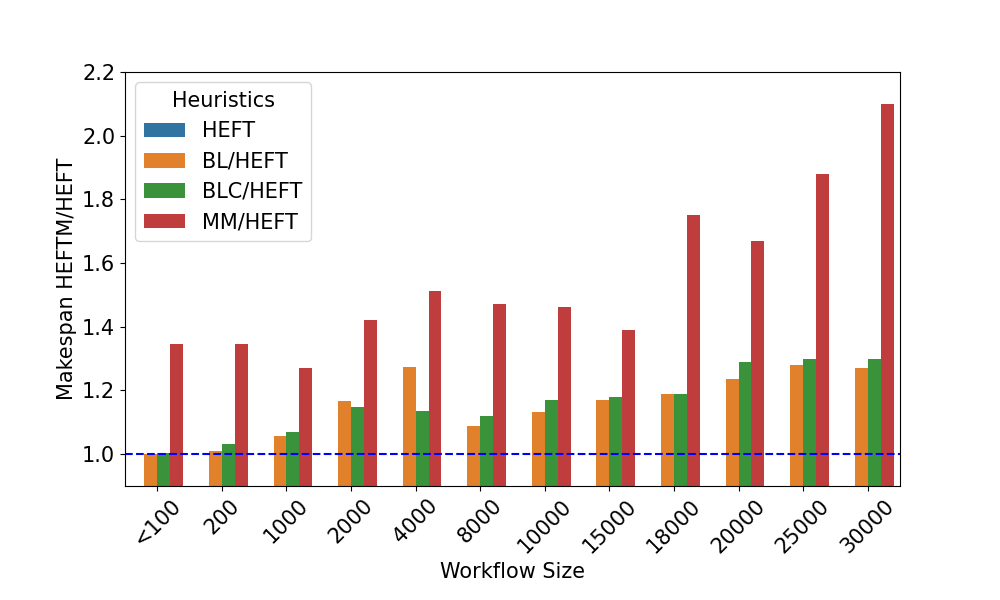}
    \caption{Relative makespans of heuristics normalized by \heft makespan, by workflow
    size, on default cluster. Smaller is better.}
    \label{fig:ms-relations-by-workflow}
%    \vspace{-0.15cm}
\end{figure}

We also report in  Figure~\ref{fig:ms-relations-by-workflow} the relative makespan found
by the memory-aware heuristics, normalized to the
makespan achieved by \heft, \new{the latter} often with an invalid over-optimistic schedule that exceeds
the memory bound.
%    Figure~\ref{fig:ms-relations-by-workflow} shows the relations of the makespans found by the heuristics to the
%    makespans found by the baseline (including invalid ones), by workflow size.
\new{The makespans found by \heftbl are $<13\%$ worse than those found by the baseline for workflows with less than \numprint{2000}
tasks, and up to $27\%$ worse for the largest workflows.
The makespans of \heftblc are very similar, being $<17\%$ worse on smallest workflows, $19$-$28\%$ worse on middle-sized,
and up to $30\%$ worse on the largest workflows.
The makespans found by \heftmm are $25$-$30\%$ worse on the smallest workflows, $40$-$70\%$ on middle-sized and up to
$210\%$ on the largest ones.} These are still very encouraging results,
in particular for \heftbl and \heftblc, since the makespans of \heft\ \new{nearly always} correspond to invalid schedules.

 %   \skug{TODO: how much worse are makespans for SUCCESSFUL hefts? presulmably almost identical}

%    \begin{figure}[tb]
%        \centering
%        \includegraphics[width=1.1\columnwidth] {images/MsRelations2}
%        \caption{Relative makespans produced by the heuristics to the makespan produced by the baseline. Smaller is better.
%        \skug{SUGESTION: remove this picture, as it duplicates Fig 3}
%        }
%
%        \label{fig:ms-relations}
%        \vspace{-0.3cm}
%    \end{figure}

%    \paragraph{Memory Usage}

\begin{figure}[tb]
    \centering
    \includegraphics[width=1\columnwidth] {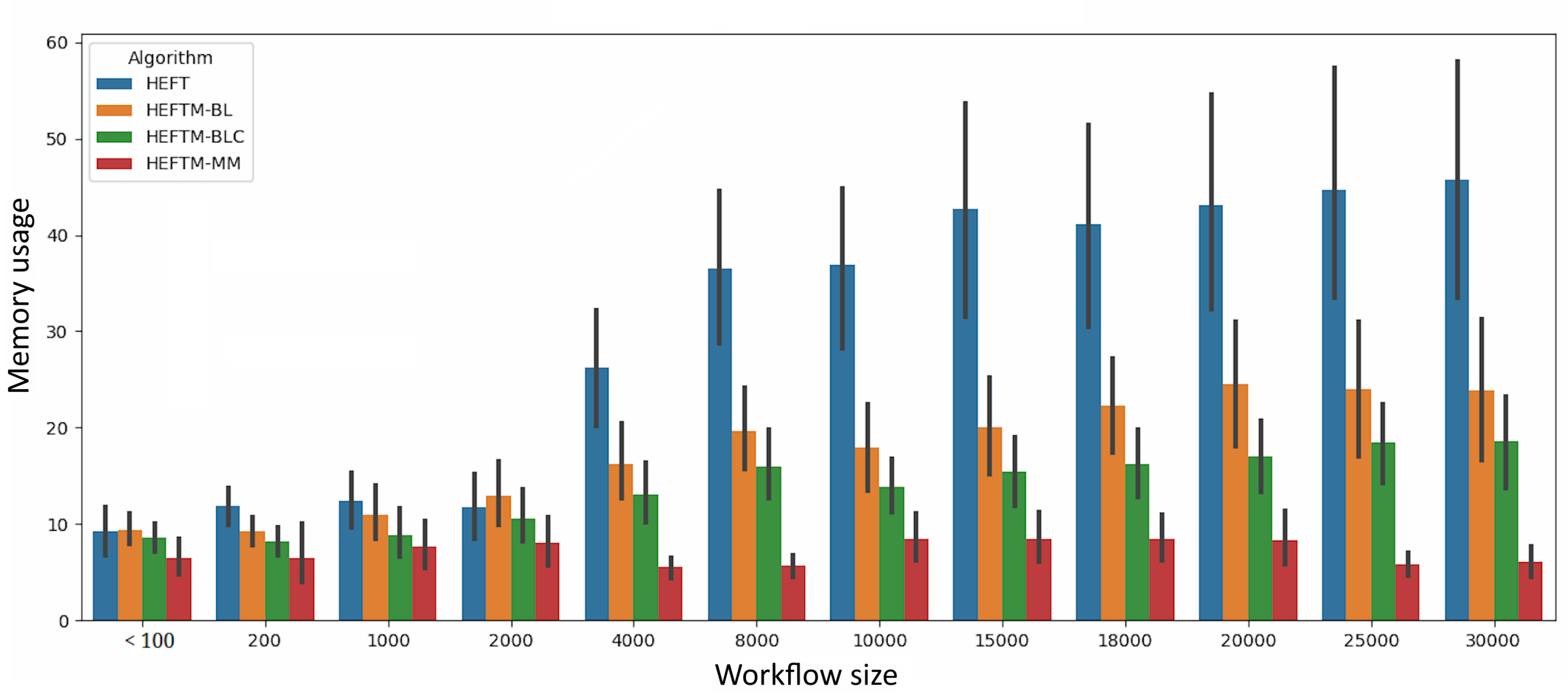}
%    \vspace{-0.6cm}
    \caption{Memory usage on default cluster, including invalid \heft schedules. }
%by algorithm by workflow size
     \label{fig:mem-usages-normal}
%    \vspace{-0.15cm}
\end{figure}

\begin{figure}[tb]
    \centering
    \includegraphics[width=1\columnwidth] {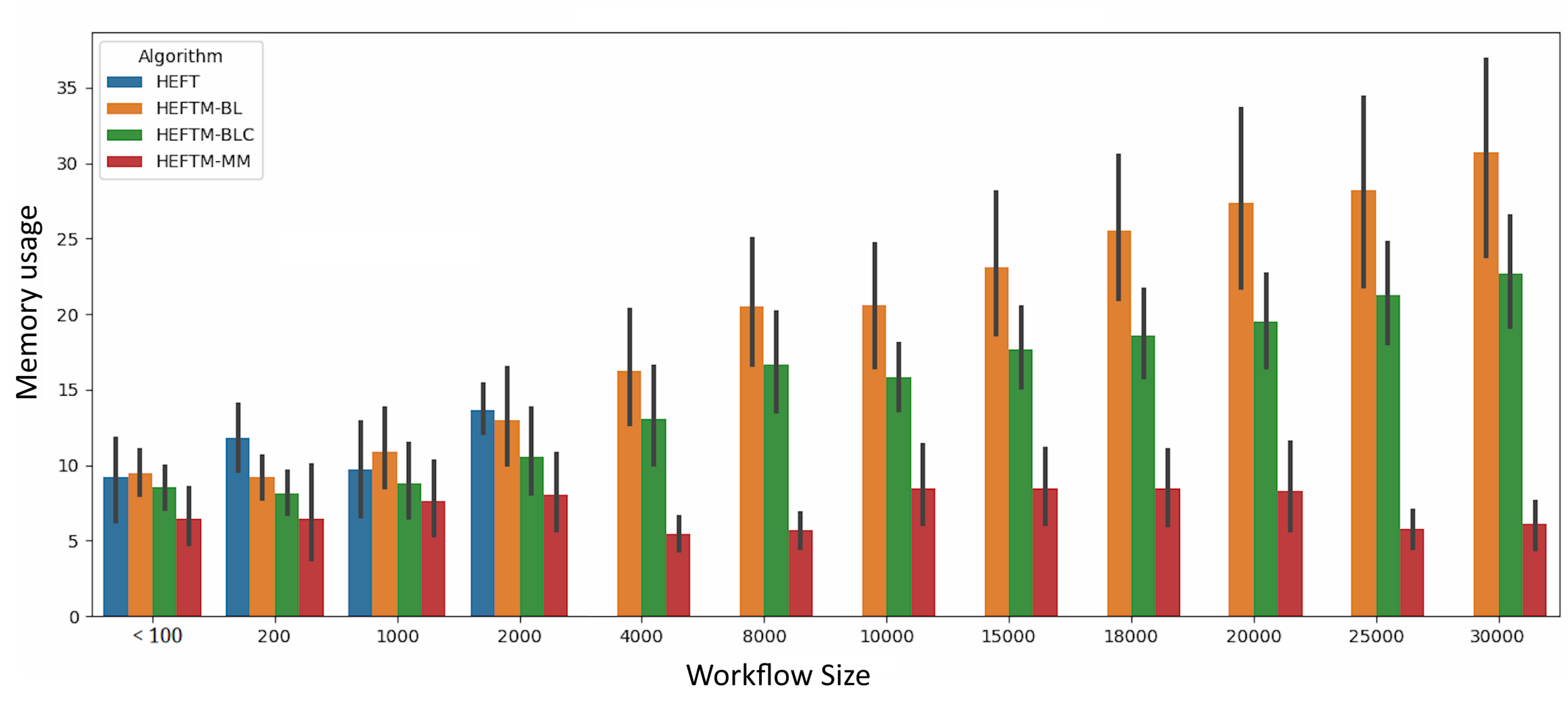}
%    \vspace{-0.6cm}
    \caption{Memory usage on default cluster, considering only valid \heft schedules. }
    \label{fig:mem-usages-onlyvalid}
%    \vspace{-0.15cm}
\end{figure}

Finally, we study the percentage of memory occupied by the schedule, which is another good indicator
of the memory usage of the heuristics and their ability to produce valid schedules.  %its important characteristic.
Figures \ref{fig:mem-usages-normal} and~\ref{fig:mem-usages-onlyvalid} show the percentage of memory
occupied on average by the schedule
produced by the different heuristics for different workflow sizes;
first on all schedules (including invalid \heft schedules) and then only on 
valid schedules (hence, no results for \heft on large workflow sizes).
%
%   Figure shows memory usages if only valid heft schedules are presented.
%    There are no valid heft scheduler from middle-sized and large workflows, so no red bars are present.

\heftmm continuously outperforms the other heuristics in terms of memory usage,
using from $46\%$ less memory on the smallest
workflows to $4$ times less memory on the largest ones with \numprint{30000} tasks.
If we consider the invalid \heft schedules, too, we see that they would require more and more memory on average,
which explains why these schedules rapidly become invalid.
This is because some assignments require more than $100\%$ of the memory (which makes them invalid).
We can assess the degree of invalidity by comparing \heft memory usage with the memory usage of \heftbl.
\heftbl differs from the baseline only in the sense that it respects the available memory on the processors.
For the largest workflows, \heft schedules require almost twice as much memory as \heftbl.

\medskip
\subsubsection{Scheduling on the memory-constrained cluster}
\label{ss:mem-constrained-cluster}
\begin{figure}[tb]
    \centering
    \includegraphics[width=1\columnwidth] {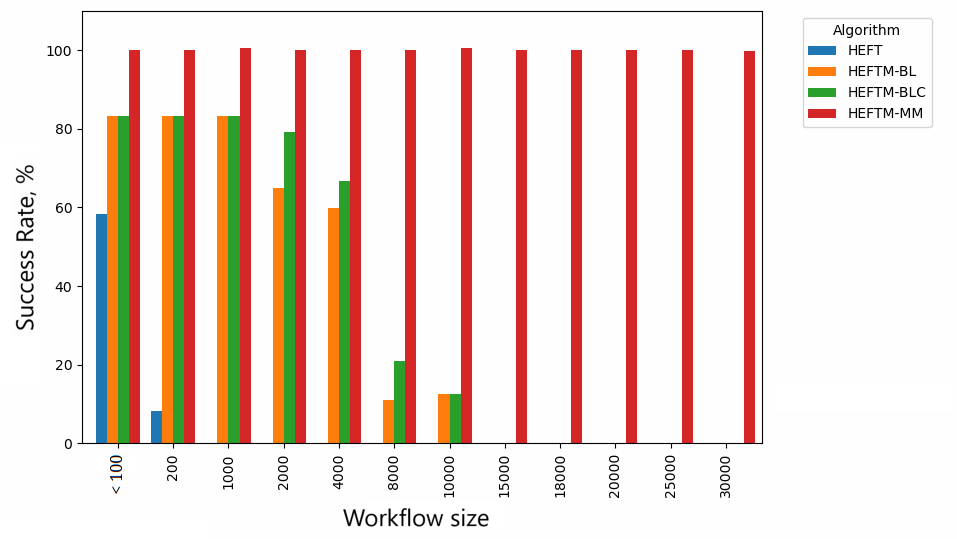}
    \caption{Success rates on the memory-constrained cluster. Higher is better. }
%        by algorithm by workflow size.
    \label{fig:success-rates-tiny}
%    \vspace{-0.15cm}
\end{figure}

On the memory-constrained cluster, \heft produces valid assignments in only $14$ out of $290$ experiments ($4.8\%$ success rate).
The successful schedules are achieved exclusively on the tiny workflows (with only two 200-task size-increased
workflow among them).
\heftbl successfully schedules $38\%$ of the workflows, \heftblc is able to schedule $49\%$ of them, while \heftmm
still schedules all of them, including even the largest ones, see Figure~\ref{fig:success-rates-tiny}.
As also observed on the default cluster, \heftmm seems to be less affected by the workflow size
than the other heuristics.

\begin{figure}[tb]
    \centering
    \vspace{-0.3cm}
    \includegraphics[width=1.03\columnwidth] {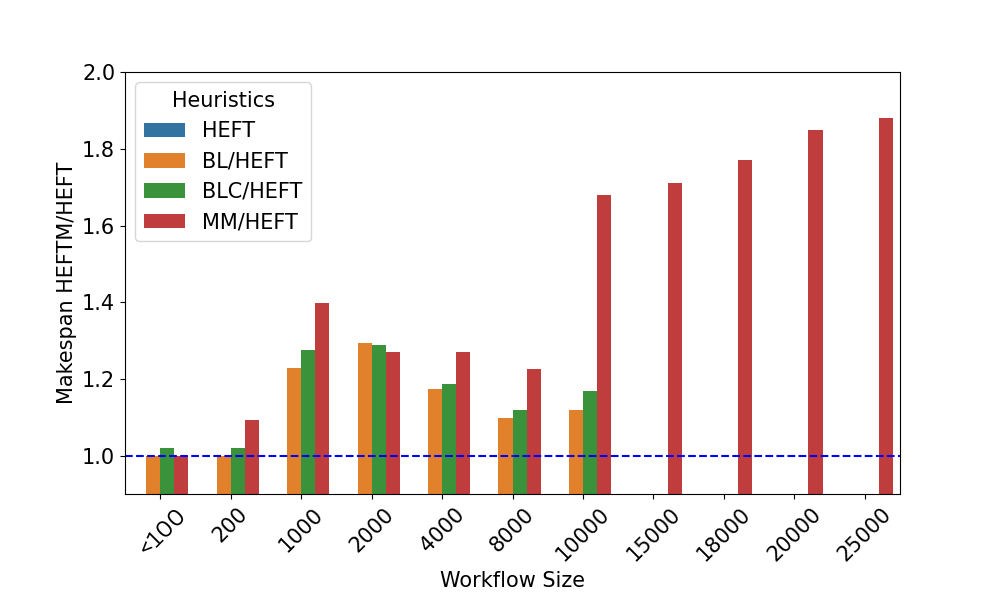}
    \caption{Relative makespans on the memory-constrained cluster.
%        produced by the heuristics to the makespan produced by the baseline, by workflow size.
    Smaller is better.}
    \label{fig:ms-relations-by-workflow-constrained}
\end{figure}

Similarly to the default cluster, we observe that the makespan of \heftmm is usually higher
than with \heft (see Fig.~\ref{fig:ms-relations-by-workflow-constrained}), but the \heft schedules are
almost all invalid. It is thus worthwhile to resort to \heftmm for large workflows
in a constrained cluster, since tasks are processed in an order that minimizes
the memory usage of schedule.
\new{Note how the relative makespan does not grow steadily with the size of the workflows.
Rather, it grows at the beginning, but then falls for the workflow sizes where scheduling becomes very challenging.
This indicates that \heft schedules become relatively worse for larger workflows, in addition to being unfeasible memory-wise.
Finally, the relative makespan for the largest workflows, the ones for which only \heftmm can produce valid schedules,
is growing larger again.
This is due to the highest degree of memory adaptation necessary to produce a valid schedule.
}

%
%    The makespans it produces are larger, but it seems to order the vertices in a similar way on both the small and the large workflows,
%    rather than resorting to more memory-consuming ordering that other heuristics and the baseline produces.
%    This makes it especially valuable for very large workflows in situations of very constrained memory in the cluster.
%
%

\begin{figure}[tb]
    \centering
    \includegraphics[width=1\columnwidth] {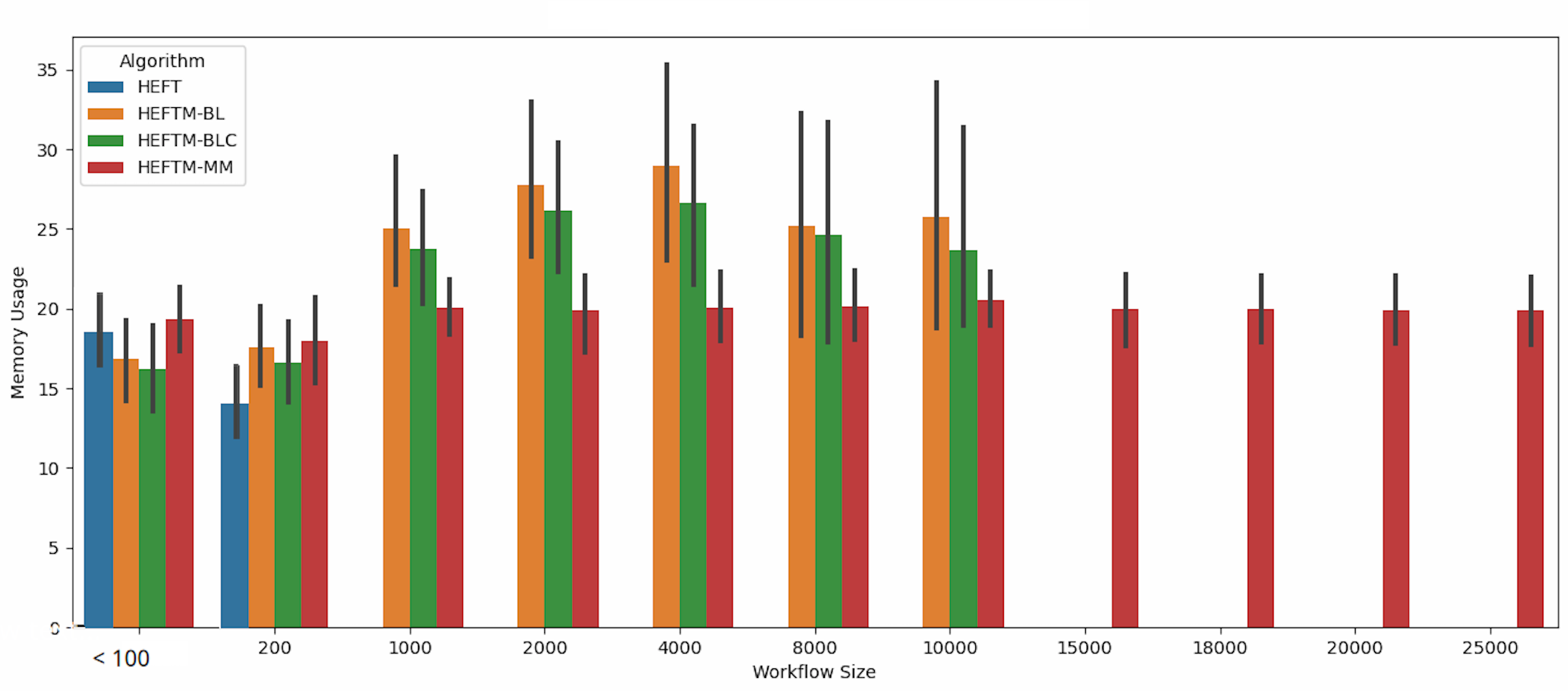}
    \caption{Memory usage on the memory-constrained  cluster. } % by different heuristics by workflow size.}
    \label{fig:memory-usage-constrained}
%    \vspace{-0.15cm}
\end{figure}

Memory usage on the constrained cluster are depicted in~Fig.~\ref{fig:memory-usage-constrained}.
\new{We plot the average memory usage in the cluster per algorithm and workflow size},
and we observe that the memory footprint of \heftmm remains constant with workflow size.

\subsection{Dynamic experiments on the memory-constrained cluster}
\label{sec.expe.dyn}
\new{In this section, we try to assess the advantage of the dynamic approach.
To this end, we compare the makespans of executions with and without recomputation.}
The makespan in case of no recomputation becomes invalid as soon as at least one task finds itself in an invalid memory size
situation -- that is, if the scheduler assumed the task's memory to be smaller than actually needed and
assigned the task to a processor with not enough memory capacity.
Due to an extremely constrained memory in this cluster, only 134 experiments out of 1160
(290 experiments for each of the three heuristics and the baseline \heft)
 \new{lead to a valid makespan when no recomputation is done}.

In the case of \heft, $14$ valid initial schedules were computed.
Out of them, $13$ remained valid \new{until the end of an execution with recomputation},
\new{i.e., the schedule is still valid when accounting for the final execution time of each task and their memory requirement.}
The same $13$ experiments also led to \new{valid schedules in case of no recomputation, i.e., when following the initial schedule without ever recomputing it}. 
Indeed, these workflows required so few resources that
\new{the deviations did not invalidate their schedules. Also, there is also nothing to be gained by rescheduling them.}
%$13$ experiments ended with a valid makespan after recomputation in case of \heft, which corresponds to $4.4\%$ success
%rate. However, there were only $14$ successful \heft experiments overall in this setup, so for the successful experiments
%only, the success rate of not recomputing is $92\%$.

In case of \heftmm, all $290$ workflows can be scheduled initially \new{(static case)} and all these schedules remain valid until the end,
\new{when new schedules are computed in a dynamic way (with recomputation). However, only}
$16$ experiments end with a successful schedule without recomputation, a rate of~$5.5\%$.

\heftblc produces $142$ valid initial schedules and keeps $141$ of them valid until the end \new{when new schedules are computed in a dynamic way.}
$50$ experiments \new{remain} successful \new{when following the original schedule} without recomputation, a rate of $35\%$ out of all successful final schedules.

\heftbl keeps $105$ schedules valid until the end out of $110$ initial \new{(static)} valid schedules.
$55$ remain valid until the end even without recomputation, roughly $50\%$ of all successes.

\heftbl and \heftblc are successful on smaller workflows and fail on larger ones. 
\new{Note that the success of the strategy without recomputation is limited to  the smallest  workflows,
while a dynamic approach is mandatory when dealing with larger workflows.}
Thus, the strategy without recomputation delivers valid makespans (independently of the algorithm) 
on $56$ original workflows with $<100$ tasks, \new{$47$ workflows with  $200$ tasks, 
$25$ workflows with $1000$ tasks, and only $6$ workflows with $2000$ tasks.
The most representative case for the dynamic experiments is hence \heftmm, since it succeeds
to achieve valid schedules in all cases thanks to recomputation, while most schedules
cannot be successful when no recomputation is done. 
}

\begin{figure}[tb]
    \centering
    \includegraphics[width=0.95\columnwidth] {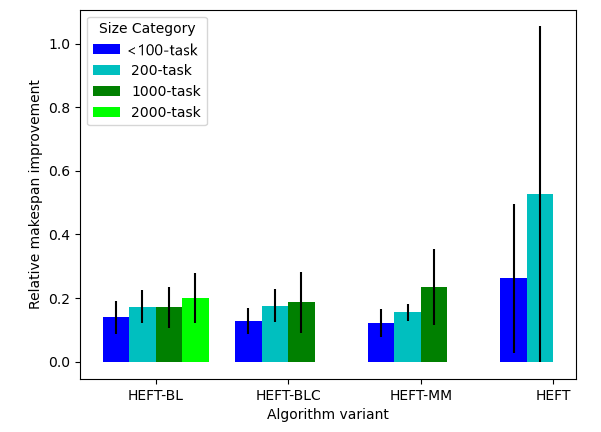}
    \caption{
    \new{Self-relative makespan improvement of \heftbl, \heftblc, \heftmm and baseline \heft against
    the execution with the same heuristic, but without recomputation. Higher is better. } }
    \label{fig:updates-ms}
%    \vspace{-0.3cm}
\end{figure}

Figure~\ref{fig:updates-ms} shows \new{the self-relative improvement in makespan when doing recomputation 
compared to no recomputation taking place} for these experiments.
With growing size of the workflow, the excess makespan of not recomputing grows -- from $13.9\%$ to $20\%$ for \heftbl,
from $12.7\%$ to $18.7\%$ on \heftblc (but there is no data for the $2000$-task workflows in this case),
% TODO: why not? SK: no (or not enough) valid solutions
$12.1\%$ to $23.5\%$ for \heftmm.
The larger variations for \heft can be explained by the small amount of data -- for instance, there are
only \new{two workflows with  $200$ tasks in this case.}

\new{
These results indicate that a meaningful comparison proves difficult due to
many invalid schedules. However, they still show that doing recomputation 
is beneficial for both the makespan and the number of valid solutions.
}

\subsection{Running times of the heuristics}
\label{sec.expe.t}
\begin{figure}[tb]
    \centering
  \hspace{-.5cm}  \includegraphics[width=1.01\columnwidth] {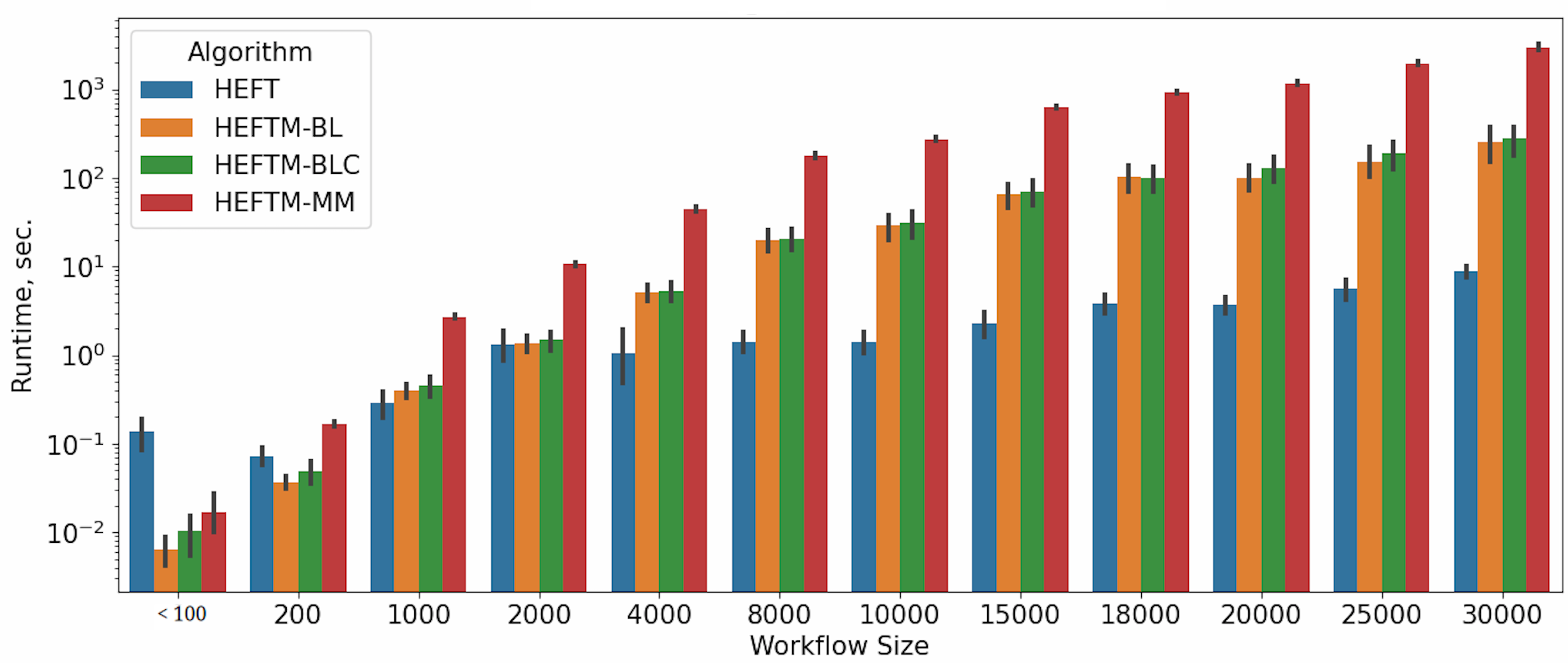}
    \caption{Average running times (in seconds) of the heuristics \wrt workflow size. The $y$-axis is logarithmic.}
    \label{fig:runtimes-log}
%    \vspace{-0.3cm}
\end{figure}

In order to answer the runtime system without holding it up for too long, the scheduler needs to 
compute a schedule reasonably quickly. %provide a fast answer.
The bottom-level-based heuristics \heftbl and \heftblc yield smaller running times than \heftmm, 
and also scale better with growing workflow sizes~(see Fig.~\ref{fig:runtimes-log}).
Their running times are similar and grow from tens of milliseconds for the smallest workflows to 
$25$-$27$ seconds for the largest workflows.
\heftmm, in turn, computes a memory-optimal traversal of the entire workflow to compute the task ranks. 
It also takes only tens of milliseconds for the smallest inputs, but requires
thousands of seconds for the large(st) workflows ($1172.7$s for \numprint{20000}-task workflows and
$2994.9$s for \numprint{30000}-task ones).
This increased running time is, however, offset by the unique $100\%$ success rate this algorithm 
obtained in our experiments when scheduling large workflows in difficult (memory-constrained) setups.

%\begin{figure}[tb]
%    \centering
%    \includegraphics[width=1.1\columnwidth] {images/runtimes-absolute}
%    \caption{Running times of the heuristics and the baseline.}
%    \label{fig:runtimes-abs}
%    \vspace{-0.3cm}
%\end{figure}

%    \subsubsection{Summary}

\bigskip
\section{Conclusion}
\label{sec:conc}

We have formalized a scheduling problem in memory-constrained environments where tasks
may exceed the memory available on a processor and resort to communication buffers
to store and communicate data between processors. In this context, we have 
designed three memory-aware HEFT-based heuristics that account for memory
constraints when scheduling workflow tasks. 

Two of these heuristics rely on a task ordering \wrt the bottom level of tasks, with the objective
of minimizing the makespan and hence scheduling critical tasks first. The third one, \heftmm,
% goes one step further and 
handles tasks in an order dictated by an efficient traversal
of the workflow in terms of memory requirements, hence reducing the memory used
by the schedule. Experimental results on a large set of workflows from real-world applications
demonstrate that the memory-aware heuristics produce valid schedules successfully. 
In the most memory-constrained setting, \heftmm succeeds to schedule even the largest workflows,
while the other heuristics return invalid schedules that exceed the memory capacity. 
This advantage of \heftmm comes at the price of worse makespans than those of \heftbl
and \heftblc. % which focus on makespan by ordering tasks by bottom level. 
As expected, the baseline \heft, which is not memory-aware, returns invalid schedules in almost
all cases, except for very small workflows. 

%Hence, \heftbl and \heftblc can schedule all workflows in a normal cluster and approximately half of them in a memory-constrained
%    one.
%    They provide a running time that scales well with the growing size of a workflow.
%
%    \heftmm can schedule even the largest workflows even on the memory-constrained cluster.
%    It is the only heuristic that has $100\%$ success rate in our experiments.
%    However, the makespans it produces are larger than those of the the other two.

Another key contribution is that we have adapted these heuristics for a dynamic setting, where exact
task parameters (execution times and memory requirements) are not known in advance. 
We have implemented a runtime system that interacts with the scheduler, providing exact
parameter values once a task arrives in the system, while only estimates are known for future tasks. 
Some preliminary experiments have been conducted in this setting and demonstrated that
it is necessary to adapt the schedule on the fly in order to avoid an execution failure because
of a memory shortage. To the best of our knowledge, this is the first study of adaptive
algorithms accounting for memory constraints. 

This work can be extended in several directions. First, the model could be refined to include
heterogeneous bandwidths. %, while we consider a homogeneous communication network. 
More importantly, it would be interesting to consider other types of variability, \egc
if new tasks (dis)appear in the workflow graph, or if there is variability in the platform, 
with processors arriving and departing. We believe that we could adapt the current approach
based on recomputing schedules on the fly, and we plan to perform a new set of experiments
to further assess the impact of dynamic scheduling techniques. 

\medskip

\subsubsection*{Acknowledgements}
This work is partially supported by Collaborative Research Center (CRC) 1404 FONDA
–- Foundations of Workflows for Large-Scale Scientific Data Analysis, which is funded
by German Research Foundation (DFG).

\balance
    \bibliographystyle{abbrv}%IEEEtran}
    \bibliography{references}

\end{document}

%% file: macros.tex
\newcommand{\iec}{\textit{i.\,e.},\xspace}

\newcommand{\egc}{\textit{e.\,g.},\xspace}
\newcommand{\eg}{\textit{e.\,g.}\xspace}

\newcommand{\etal}{\textit{et al.}\xspace}
\renewcommand{\etal}{{et al.}\xspace}

\newcommand{\wrt}{w.\,r.\,t.\xspace}

\newcommand{\parent}{\operatorname{parent}\xspace}